\documentclass[aps,prb,preprint,graphicx,amsbsy,amssymb]{revtex4}

\usepackage{graphicx}
\usepackage{amsmath}

\begin{document}
\draft
\title{Power-law Conductivity inside the Mott gap: application to $\kappa-(BEDT-TTF)_2Cu_2(CN)_3$}
\author{Tai-Kai Ng$^1$ and Patrick A. Lee$^2$}
\address{
$^1$ Department of Physics, Hong Kong University of Science and
Technology, Clear Water Bay Road, Kowloon, Hong Kong. \\
 $^2$ Department of Physics, Massachusetts Institute of Technology,
Cambridge, Massachusetts 02139, USA }
\date{ \today }

\begin{abstract}
  The charge dynamics of spin-liquid states described by $U(1)$ gauge theory coupling to fermionic spinons
  is discussed in this paper. We find that the gapless spinons give rise to a power law optical
  conductivity inside the charge gap. The theory is applied to explain the unusual optical conductivity observed
  recently in the organic compound $\kappa-(BEDT-TTF)_2Cu_2(CN)_3$. We also propose an optical experiment to
  search for the in-gap excitations in the Kagome spin liquid insulator.

\end{abstract}


\maketitle

  Recent work has shown that the organic compound $\kappa-(BEDT-TTF)_2Cu_2(CN)_3$\cite{e1,e2,e3} and the
  spin-1/2 Kagom$\grave{e}$ system $ZnCu_3(OH)_6Cl_2$\cite{e4,e5,e6} hold great promise as the first two examples
  of spin liquid states realized in dimensions great than one\cite{t1,t2,t3}. Spin liquid states are Mott
  insulators with an odd number of spin ${1\over 2}$ per unit cell which shows no long-range magnetic order. They
  are proposed to exist in systems either in the vicinity of the Mott transition\cite{t1} or with frustrated
  lattice structures. In both cases the system can be modelled by an appropriate Hubbard model with on-site
  repulsion $U$ and hopping integral $t$ at half filling. For large enough $U$ compared with $t$, charge
  excitations are gapped and the system is a spin liquid if long-range order is absent in the spin sector.

  The two recently discovered systems are believed to be two-dimensional spin liquids. In the case of
  $\kappa-(BEDT-TTF)_2Cu_2(CN)_3$ the system is described by a Hubbard model on a triangular lattice. Since the system can be driven metallic (indeed superconducting) under pressure, it is believed
  that $U$ is not very large compared with $t$
  and the insulator is near the Mott transition\cite{t1,t2}. In this case the charge excitations acquire a gap,
   and it is proposed that spin-charge separation occurs and spin ${1\over 2}$ excitations (spinons)
  form a Fermi surface\cite{t1,t2}. In the case of $ZnCu_3(OH)_6Cl_2$, it is believed that $U>>t$ and the spin
  dynamics is described
  by the antiferromagnetic Heisenberg model. The frustrated Kagom$\grave{e}$ lattice gives rise to a spin liquid
  state with Dirac fermions excitation spectrum\cite{t3}. A unique feature of these spin liquid states is that
  the spin excitations are always coupled to internal U(1)
  gauge fields representing spin-chirality fluctuations\cite{t1,t2,t3} in the spin systems.

  It is often thought the Mott insulator should be fully gaped in its optical (charge) responses.  Furthermore,
  the spinons are considered to be neutral and do not absorb electromagnetic radiation.  Here we point out that
  due to coupling with the gauge field, the spinons do contribute to optical conductivity, yielding a power law
  absorption at low frequencies.  This may explain some puzzling experimental observations recently reported in the
  organics\cite{oc}.

    The dynamics of the spin liquid states can be studied in a slave-rotor representation of Hubbard
  models\cite{t1} with the appropriate lattice structures. In this presentation the electron operator is
  represented as $c(c^+)_{i\sigma}=f(f^+)_{i\sigma}e^{-(+)i\theta}$, where $f(f^+)_{i\sigma}$ is the spin
  annihilation(creation) operator and $e^{-(+)i\theta}$ lowers(raises) the charge by one. The total charge operator
  $\rho_i=\sum_{\sigma}f^+_{i\sigma}f_{i\sigma}-1$ is the conjugate variable to $\theta$ in this representation.

    After making a mean-field approximation, the low energy effective action of the system can be written in
  terms of $\theta$ and $f(f^+)$ fields separately, $L^{1(2)}=L_c+L_s^{1(2)}$, where $L_c$ representing the
  charge dynamics and $L_s$ represents the spin dynamics of the system. $L_c$ is described by the
  strong coupling phase of a quantum $x-y$ model\cite{fg},
 \begin{subequations}
 \label{lagrangian}
 \begin{equation}
 \label{l1}
 L_c\sim\sum_i{1\over U}|(\partial_t-i(a_0+A_0))\theta_i|^2-t_{eff}\sum_{<i,j>}cos(\theta_i-\theta_j-(\vec{a}_{ij}
 +A_{ij}))
 \end{equation}
 coupling to internal gauge fields $(a_0,\vec{a})$, where $t_{eff}\sim \alpha t$ with $\alpha<1$ being a numerical
 factor determined self-consistently from the mean-field equation, $(A_0,\vec{A})$ represents the real
 electromagnetic field coupling to the system and
 \begin{equation}
 \label{l21}
 L_s^{(1)}=\sum_{\sigma}\left(f_{\sigma}^+(\partial_t-ia_0-\mu_f)f_{\sigma}-
    {1\over2m_s}f^*_{\sigma}(-i\nabla-\vec{a})^2f_{\sigma}\right)
 \end{equation}
  in the case of $\kappa-(BEDT-TTF)_2Cu_2(CN)_3$ which is believed to possess a spinon Fermi surface.
  $\mu_f$ is the chemical potential, $m_s^{-1}$ is expected to be of order of the exchange $J\sim t^2/U$. In the case of
  $ZnCu_3(OH)_6Cl_2$ where spinons have a Dirac fermion spectrum,

 \begin{equation}
 \label{l22}
  L_s^{(2)}=\sum_{\mu\sigma}\left(\bar{\psi}_{+\sigma}(\partial_{\mu}-i(a_{\mu}+A_{\mu})\tau_{\mu}\psi_{+\sigma}+
    \bar{\psi}_{-\sigma}(\partial_{\mu}-(a_{\mu}+A_{\mu})\tau_{\mu}\psi_{-\sigma}\right),
 \end{equation}
 \end{subequations}
  where $\mu=0,1,2$ and $\tau_{\mu}$ are Pauli matrices. The two-component Dirac spinor fields $\psi_{\pm\sigma}$
  describe two inequivalent Dirac nodes in the spinon spectrum\cite{t3}.

   Effects of disorder and phonons can also be included in the actions. Their contributions can be included by an
   adding a term
   \[
   L'=\sum_{\vec{p},\vec{q}}\left(V(q)c^+_{\vec{p}+\vec{q}\sigma}c_{\vec{p}\sigma}
    +M(q)c^+_{\vec{p}+\vec{q}\sigma}c_{\vec{p}\sigma}(b_{\vec{q}}+b^+_{-\vec{q}})+
    b^+_{\vec{q}}(\partial_0-\omega_{\vec{q}})b_{\vec{q}}\right)
    \]
  to $L_s^{(1)}$, where $V(q)$ is a disordered potential and $b(b^+)_{\vec{q}}$ are phonon annihilation (creation) operators
  with momentum $\vec{q}$. $M(q)$ is the electron-phonon coupling and $\omega_{\vec{q}}$ is the phonon
  dispersion. A corresponding term can also be added to $L_s^{(2)}$ for Dirac fermions.

   The thermodynamic and magnetic properties of the above systems have been studied in several previous
  papers\cite{t1,t2,t3,nave}. We shall concentrate on the charge dynamics of these spin liquid states here. We
  assume a Mott insulator state with no broken symmetry and with isotropy in space.  The current response
  function is given by the conductivity, which can be decomposed into longitudinal and transverse parts
  $\sigma_\parallel$ and $\sigma_\perp$.  For a $U(1)$ spin liquid, the Ioffe-Larkin composition rule\cite{Ioffe}
  relates the physical $\sigma$ to the response function of the spin and charge
  components,

  \begin{equation}
  \label{il1}
  \sigma_\perp(q,\omega) =  \left(
  \sigma_{s\perp}^{-1}(q,\omega) +
  \sigma_{c\perp}^{-1}(q,\omega)
  \right)^{-1}
  \end{equation}
  and similarly for $\sigma_\parallel$.  Here $\sigma_s$ and
  $\sigma_c$ are given by the {\em proper} response functions of the spin and charge (represented by $\theta$)
  fields appearing in the action $L_s$ and $L_c$, respectively. The {\em proper} response
  functions represent sum of all diagrams which cannot be separated into two parts by cutting one
  interaction line associated with either the real or internal gauge field, and represents the current response of
  the charges and spinons to the potential $\vec{a}+\vec{A}$ and $\vec{a},$\cite{Ioffe,nl}
  respectively. Notice that both the phonon and impurity contributions can be included in the definition
  of the {\em proper} response functions. The origin of the Ioffe-Larkin rule is that an external $\vec{A}$ field
  induces a nonzero $\vec{a}$ field which is needed to enforce the constraint $j_{c\mu} + j_{s\mu} = 0$.\cite{nl}
  Thus even though the $\vec{A}$ field couples only to the $\theta$ field, the induced $\vec{a}$ field indirectly
  couples to the gapless spinons.

  We parametrize the longitudinal response of the charge field by a dielectric constant $\varepsilon_c$  and
  ignore the analytic correction in $q^2, \omega^2$ for small $q$ and $\omega$.  Then

  \begin{equation}
  \label{ep}
  \varepsilon_c = 1 + {4\pi i  \sigma_{c\parallel}\over \omega}.
  \end{equation}

 We expect $\varepsilon_c -1$ to decrease with increasing charge gap.  Furthermore, for small $q$ there is no distinction
 between longitudinal and transverse response in an insulator.  Using\ (\ref{ep}) for both, we find using Eq.\
 (\ref{il1}),

 \begin{equation}
 \label{sig}
 \sigma_{\parallel(\perp)} (q,\omega) =
 {\omega\sigma_{s,\parallel(\perp)}(q,\omega)
  \over
 \omega + i \left( {4\pi \over \varepsilon_{c}-1} \right)  \sigma_
 {s,\parallel(\perp)}
 (q,\omega)}.
\end{equation}
 We should point out that the replacement of the charge response by a dielectric constant is not as innocent as it
 appears.  This step should be considered in the spirit of random phase approximation and justified using a
 $1\over N$ expansion.  The concern is the existence of Feynman diagrams involving multiple gauge field lines
 going across.  In the language of proper response function, these become part of the charge vertex which couples
 to the external gauge field.  Since the gauge field carries gapless excitations, the approximation of this
 vertex by a dielectric constant is not strictly correct except as leading order in $1\over N$\cite{Kim}.

  Now we consider the optical conductivity given by $\sigma_\perp(q=0, \omega)$.  In this limit there is no
 distinction between longitudinal and transverse and we can drop the $\perp$ subscript. The spinon conductivity
 is expected to be metallic-like. We can safely assume $Re[\sigma_s(0,\omega)]>>\omega$ and
 $Im[\sigma_s]<<Re[\sigma_s]$ for small $\omega$, and we obtain from Eq.\ (\ref{sig})

\begin{equation}
\label{sig1}
 Re[\sigma (\omega)] = \omega^2 \left(
{\varepsilon_c-1 \over 4\pi} \right)^2 {1\over Re[\sigma_s
(\omega)]}.
\end{equation}
 Note that $Re\sigma (\omega) = 0$ for $\omega = 0$ as expected for an insulator, but we find contribution inside
 the gap for small $\omega$.  First we consider the case when disorder scattering of the spin is weak.  Then
 $\sigma_s(\omega) = ne^2 \tau (\omega,T)/m_s$.  The dominant contribution to $\tau^{-1}$ is inelastic scattering
 due to the gauge field,\cite{nl} which is given by ${1\over \tau} \sim [max (\hbar\omega,k_BT)]^{4/3}$.  For
 $\hbar \omega > \hbar/\tau_0 , kT$ where $\tau_0$ is the elastic scattering time, we find

\begin{equation}
Re [\sigma(\omega)] = \omega^{3.33} \left( {\varepsilon_c-1 \over
4\pi} \right)^2
 {m_s \over n}
\end{equation}
in qualitative agreement with what is observed experimentally in $\kappa-(BEDT-TTF)_2Cu_2(CN)_3$\cite{oc}. Our
   theory also predicts that $Re[\sigma(\omega)]$ crossover to $\sim\omega^2$ at small enough $\omega$.
   The above results are strongly modified if localization effect is important and $\sigma(\omega)$ vanishes at
   $\omega\rightarrow0$ faster than $\omega$. In this case
   \[
   \sigma(\omega)\sim \sigma_s(\omega)  \]
   will show similar behavior as observed in usual strongly disordered metals.

     The above analysis can be generalized straightforwardly to the Kagom$\grave{e}$ system $ZnCu_3(OH)_6Cl_2$
   which is believed to be a spin liquid with Dirac fermion excitation spectrum $\omega = \bar{c}q$. The only
   difference is that the ``proper" response functions should be replaced by the
   corresponding functions for Dirac fermions. In this case $\sigma_s(q,\omega)$ has the universal form
   \begin{equation}
   \label{jjdirac}
    \sigma_s(q,\omega)\sim{e^2\over8}{(\bar{c}^2q^2-\omega^2)^{{1+\beta\over2}}\over i\omega}
   \end{equation}
   where $\beta=0$ for non-interacting Dirac fermions and is an unknown exponent in the presence of gauge field
   interaction. Putting $\sigma_s(0,\omega)$ into Eq.\ (\ref{sig}), we predict
   $Re[\sigma(\omega)] \propto \omega^{2-\beta}$ for $\beta<1$ and $Re[\sigma(\omega)] \propto \omega^{\beta}$ for
   $\beta>1$ and we see that the optical conductivity probes directly the unknown exponent $\beta$. Since the
   Kagome system is deep in the Mott insulator regime, the observation of power law conductivity inside the Mott
   gap strong than $\omega^4$ (see Eq.\ (\ref{re}) below) will be strong evidence for the existence of gapless
   spinons.

        We shall now study the general dielectric response $\varepsilon (q,\omega)$ of
   $\kappa-(BEDT-TTF)_2Cu_2(CN)_3$, which is believed to possess a spinon Fermi surface in more detail. We shall
   assume that the residual interactions are weak enough so that the spinons are in a
   Fermi liquid state. The proper density-density response function of the system is $\chi_d(q,\omega)$
  which represents the sum of all polarization diagrams which cannot be separated into two parts by cutting
   one Coulomb interaction line associated with the real electromagnetic
   field\cite{Ioffe,Kim} The dielectric function of the spin liquid is therefore
   \[
   \varepsilon(q,\omega)=1-v_e(q)\chi_d(q,\omega)
   \]
   where $v_e(q)=4\pi e^2/q^2$ is the real Coulomb interaction. We assume here that the (3D) system is a sum of
   layers of spin liquid here.

 Charge conservation gives $\chi_d = ({q^2/\omega^2})\chi_\parallel$, where $\chi_\parallel$ is the longitudinal
 current-current response function which is in turn given by $\sigma_\parallel = e^2\chi_\parallel/i\omega$.
 Combining these relations we obtain the usual formula

 \begin{equation}
 \label{ep2}
 \varepsilon(q,\omega) = 1 + 4\pi  i \sigma_\parallel
 (q,\omega)/\omega
 \end{equation}
 where $\sigma_\parallel$ is related to $\sigma_{s\parallel}$ by Eq.\ (\ref{sig}). In the absence of scattering,
 we expect the density-density response function to be

 \begin{equation}
 \label{chi}
 \chi_{ds} = {dn \over d\mu} + {i \gamma \omega \over v_Fq}
 \end{equation}
 where $\gamma$ is the quasiparticle density of states at the Fermi level and $v_F$ is the Fermi velocity.
 Eq.\ (\ref{chi}) is valid in Fermi liquid theory and has been shown to remain applicable for small $q,w$ when gauge
 fluctuations are treated to two loop order.\cite{Kim}  Particle conservation again allows us to write
 $\sigma_{s\parallel} =i\omega \chi_{ds}/q^2$. Combining these results we find

 \begin{eqnarray}
 \sigma_\parallel (q,\omega) &=& \left( {\varepsilon_c-1\over 4\pi}
 \right) {\omega \over i} \left( 1 - {i\omega \over
 \sigma_{s\parallel}} \left( {\varepsilon_c-1\over 4\pi} \right)
 \right)^{-1}
 \nonumber \\
 &=& \left({\varepsilon_c-1 \over 4\pi}\right) {\omega\over i}
 \left( 1 - {q^2 \over \chi_{ds}} \left( {\varepsilon_c-1 \over
 4\pi} \right) \right)^{-1}.
 \end{eqnarray}
 Using Eq.\ (\ref{ep2}), we obtain at small $q$
 \begin{equation}
 \varepsilon(q,\omega) = \varepsilon_c + {\left(
 (\varepsilon_c-1)^2/4\pi \right) q^2 \over {dn \over d\mu} +
 {i\gamma \omega \over v_Fq} }.
 \end{equation}

The static dielectric constant is given by the charge part
$\varepsilon_c$ and the full dielectric function is in principle
measurable by electron diffraction.

      Phonons have small effects on the above results. It only modifies the interaction parameter $\gamma$ and
    renormalizes the compressibility $\partial n/\partial\mu$. The effect of disorder can be included by modifying
    $\chi_{ds}(q,\omega)$ into a diffusive form
    ${dn \over d\mu} {Dq^2 \over Dq^2 + i\omega}$
    if localization effect is not important\cite{lreview}. In this case, we obtain

 \begin{equation}
 \label{ep3}
 \varepsilon(q,\omega) = \varepsilon_c + {(\varepsilon_c-1)^2(Dq^2 + i\omega)
\over 4\pi \sigma_{s,\parallel} }
\end{equation}
 where $D$ is the spinon diffusion constant and $\sigma_{s,\parallel} = e^2{dn \over d\mu} D$. For $q = 0$,
 Eq.\ (\ref{ep3}) is consistent with the AC conductivity given by Eq.\ (\ref{sig1}) as expected.

      It should be emphasized that the coupling of density and current responses to spin excitations exists rather
   generally in insulators and does not rely on existence of a spin-liquid state. Assuming that the electronic
   properties of the insulator is described by a Lagrangian with an one-particle term and an effective
   electron-spin coupling of form
   \[
   L'=\vec{S}.(\psi^+\vec{\tau}\psi), \]
   where $\psi=(c_{\uparrow},c_{\downarrow})$ is a 2-component spinor where $c_{\sigma}$'s are electron operators
   and $\vec{S}$ is an effective spin operator, the leading order coupling terms between spins and density/current
   fluctuations can be derived and are represented in the Feynman Diagram shown in Fig. (1a), where
   the solid lines are electron propagators.
   \begin{figure}
   \includegraphics[width=4.5cm, angle=0]{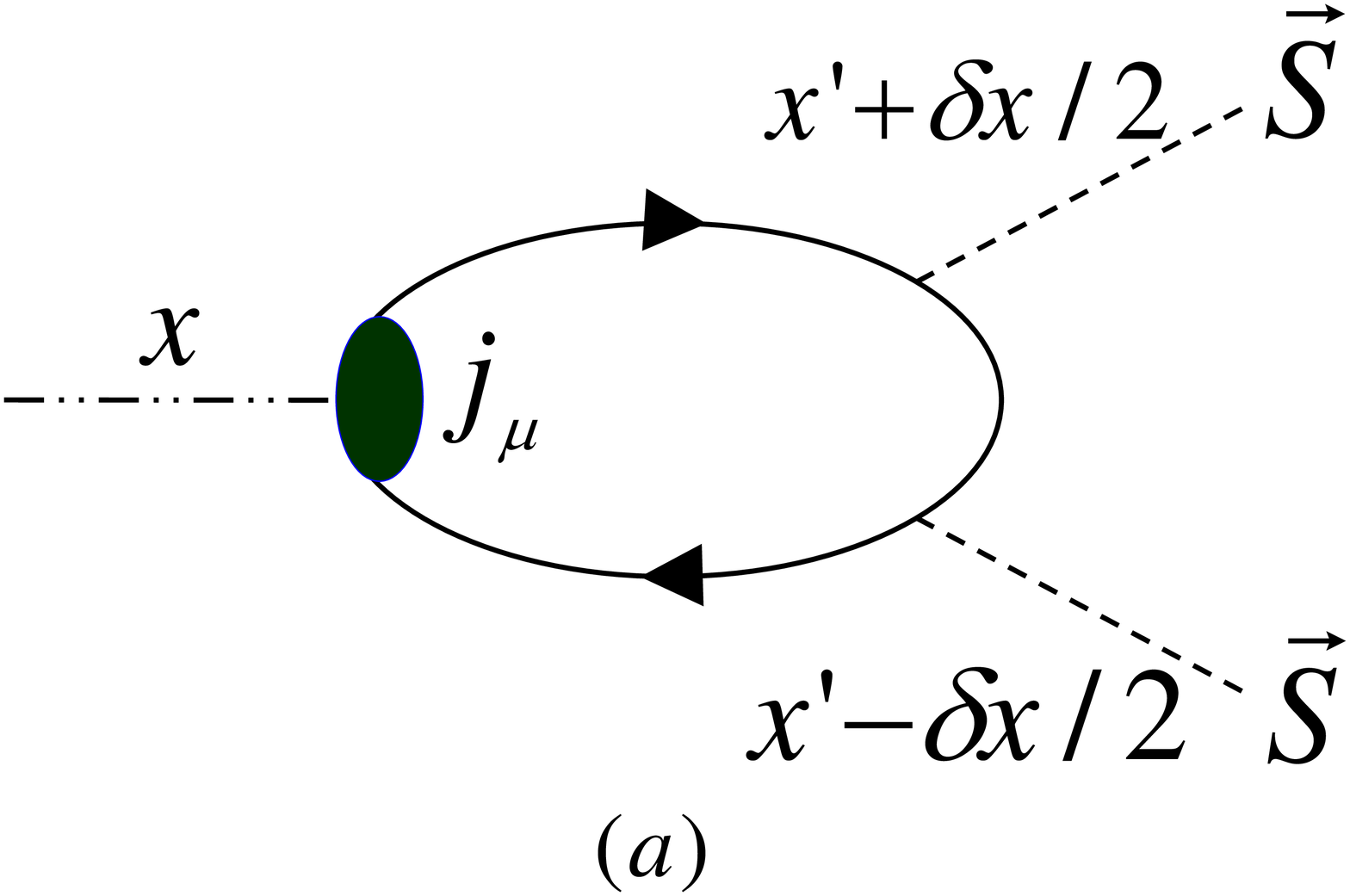}
   \includegraphics[width=4.5cm, angle=0]{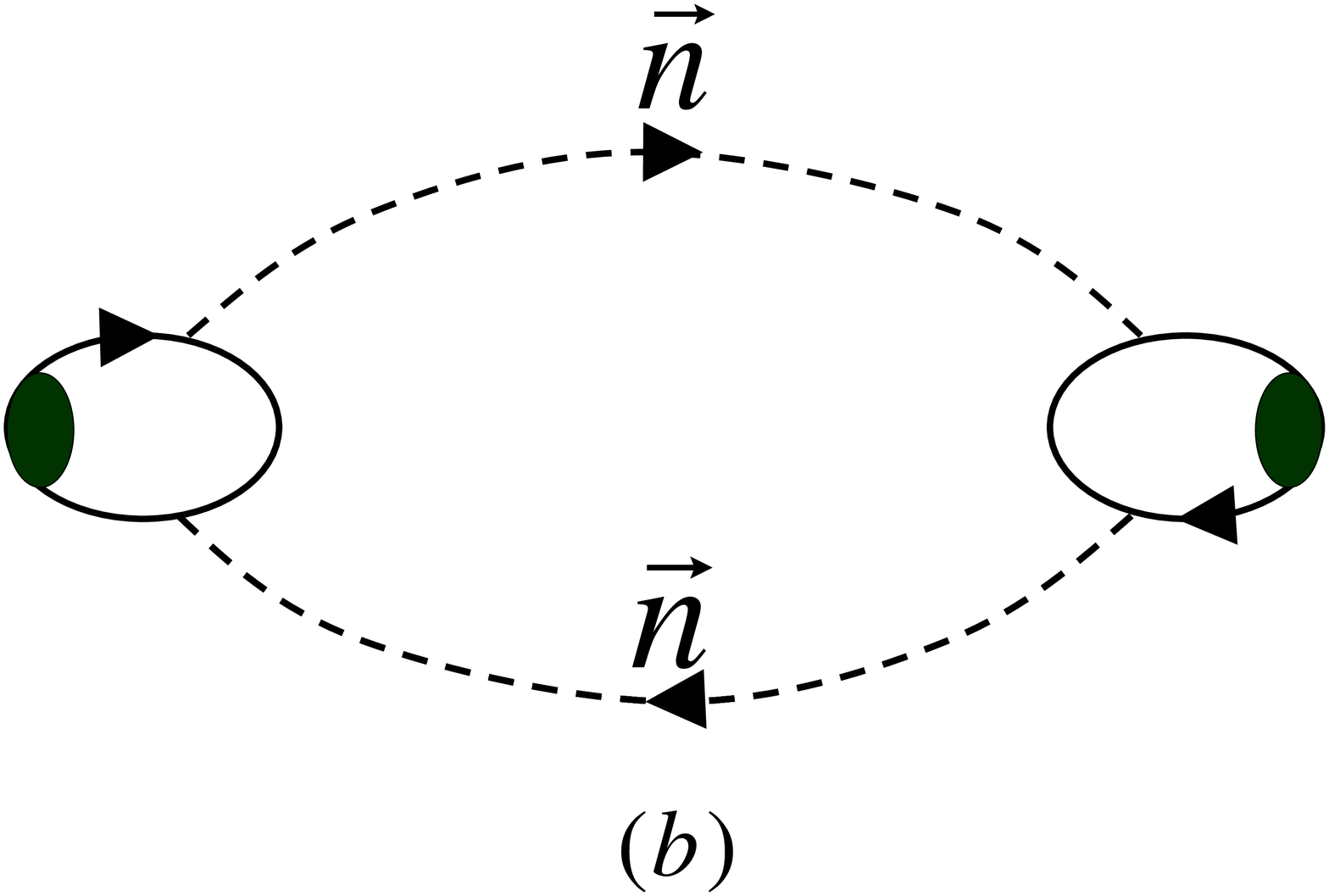}
   \caption{\label{Fig.1}
    (1a) Leading order Feynman diagram representing coupling between spin and density/current fluctuations. Solid lines
    represent electron Green's functions. There is another diagram where electron lines reverse in direction.
    (1b) Corresponding Feynman diagram representing correction to proper density-density response
    function.}
   \end{figure}
   In real space-time, the diagrams are represented by expressions of form
   \begin{eqnarray}
   \label{g1}
   \Gamma_{\mu}(x,x'_-,x'_+;\vec{S}) & = & \sum_{\sigma\sigma'\nu\nu '}G_{\sigma}(x-x'_-)\hat{j}_{\mu}(x)
   G_{\sigma}(x'_+-x)S^{\nu}(x'_-)\tau_{\sigma\sigma '}^{\nu}G_{\sigma'}(\delta x)
   S^{\nu '}(x'_+)\tau_{\sigma '\sigma}^{\nu '}
   \\  \nonumber
   & = & \sum_{\sigma}G_{\sigma}(x-x'_-)\hat{j}_{\mu}(x)G_{\sigma}(x'_+-x)\left[G_{-\sigma}(\delta x)(S^x(x'_-)S^x(x'_+)
   +S^y(x'_-)S^y(x'_+))\right.
   \\  \nonumber
   & & +\left.G_{\sigma}(\delta x)S^z(x'_-)S^z(x'_+)-i(\sigma)G_{-\sigma}(\delta
   x)\left(S^x(x'_-)S^y(x'_+)-S^y(x'_-)S^x(x'_+)\right)\right]
   \end{eqnarray}
   where $\hat{j}_{\mu} (\mu=0,1,2)$ is the electron current operator and $x=(\vec{x},t)$, $x'_{-(+)}=x'-(+)
   \delta x/2$. Assuming that the electrons have a gapped spectrum (insulator), the corresponding Green's
   function $G_{\sigma}(x)$ is short-ranged and the contributions mainly come from small $\delta x$ region. Therefore we
   can expand $G_{\sigma}(x'\pm\delta x/2-x)\sim G_{\sigma}(x'-x)\pm(\delta x/2)\partial_xG_{\sigma}(x'-x)+..$,
   $S^{\nu}(x'\pm\delta x/2)\sim S^{\nu}(x')\pm(\delta x/2).\partial_{x'}S^{\nu}+..$, etc. in Eq.\ (\ref{g1}) to
   derive the leading order spin-density(current) coupling terms in the insulating state in the continuum
   limit. A corresponding expansion for metallic ferromagnetic states has been developed previously\cite{nayak}.
   By keeping two sites per unit cell, this procedure can be extended to derive the correction to optical
   conductivity in the antiferromagnetically ordered state in the Hubbard
   model, which is a competing state to the spin-liquid state observed in the organic compound
   $\kappa-(BEDT-TTF)_2Cu_2(CN)_3$\cite{oc}. In this case, $G(x)\rightarrow  G^{ab}(x)$ and
   $\vec{S}(x)\rightarrow\vec{S}^{a}(x)=\vec{m}(\vec{x})+(-1)^a\vec{n}(\vec{x})$, where $a,b=A,B$ are sublattice
   indices. $\vec{m}$ and $\vec{n}$ represent magnetization and staggered magnetization fluctuations, respectively.
   The low energy contribution to optical conductivity is dominated by coupling of density fluctuations to two
   spinwave process represented by coupling to $\vec{n}$ fields. After some algebra, we obtain in the small
   wave-vector limit,
    \begin{subequations}
   \label{af}
   \begin{equation}
   \label{gamma}
   \Gamma_0(q,\omega;q',\Omega;q-q',\omega-\Omega;\vec{S})\sim\omega(\vec{q}.\vec{q}')
    \vec{n}(\vec{q}',\Omega).\vec{n}(\vec{q}-\vec{q}',\omega-\Omega).
   \end{equation}

     We have assumed that the antiferromagnetic state is described  by usual mean-field theory with non-zero
     staggered magnetization $<m>$. The corresponding correction to proper density density response function
     (Fig.(1b)) is given by
   \begin{equation}
   \label{cor}
   \delta\chi_d(0,\omega)\sim{1\over V\beta}\sum_{q'\Omega}{|\Gamma_0(0,\omega;q',\Omega;-q',\omega-\Omega;\vec{S})|^2\over
   (\Omega^2-c_m^2q'^2)((\omega-\Omega)^2-c_m^2q')^2},
   \end{equation}
   where $c_m\sim U<m>$ is the spinwave velocity derived from the mean-field theory.
   Evaluating the integral, we find that the correction to optical conductivity is
   \begin{equation}
   \label{re}
   \delta\sigma(\omega)\sim e^2({\omega\over c_m})^{d+2},
   \end{equation}
    \end{subequations}
     for $\omega<<U<m>$, where $d$ is the dimension. We have assumed $t\sim U$ in our calculation. Notice that the
     optically conductivity is {\em enhanced} in the spin-liquid state compared with the antiferromagnetically
     ordered state, in agreement with what is observed experimentally\cite{oc}.

     In conclusion, we have shown that gapless spinons in a spin liquid state gives rise to a power-law optical
   absorption inside the Mott gap which is larger than that expected for two spin wave absorption in a Neel ordered
   insulator. Recent experiment has reported the surprising finding that the low temperature optical absorption in
   $\kappa-(BEDT-TTF)_2Cu_2(CN)_3$ is larger than another compound $\kappa-(BEDT-TTF)_2Cu[N(CN)_2]Cl$ which exhibit Neel ordering
   but is "closer" to the Mott transition in that it has a smaller Mott gap\cite{oc}. Our result gives a natural
   explanation of this puzzle. We believe that power-law absoption, especially if it can be observed in a large
   gap insulator such as the Kagome system, is strong evidence for the existence of gapless spinons and gauge
   fields.

  \acknowledgements
  T.K. Ng acknowledge support from HKUGC through grant CA05/06.SC04.  P.A. Lee acknowledges support by NSF DMR--0517222.

\end{document}